\documentclass[12pt,a4paper,final]{article}

\usepackage{a4wide}
\usepackage{amssymb}
\usepackage[english]{babel}
\usepackage{color}
\usepackage[colorlinks=false,%
            pdfkeywords={OCaml, Just In Time compilation, machine code generation},%
            pdftitle={OCamlJIT 2.0 - Faster Objective Caml},%
            pdfauthor={Benedikt Meurer},%
            pdfsubject={},%
            pdfdisplaydoctitle=true]{hyperref}
\usepackage{tikz}
\usepackage{varwidth}

\usetikzlibrary{arrows}
\usetikzlibrary{trees}

\begin{document}

\title{%
  OCamlJIT 2.0 - Faster Objective Caml
}
\author{%
  Benedikt Meurer\\
  Compilerbau und Softwareanalyse\\
  Universit\"at Siegen\\
  D-57068 Siegen, Germany\\
  \url{meurer@informatik.uni-siegen.de}
}
\date{}
\maketitle
\begin{abstract}
  This paper presents the current state of an ongoing research
  project to improve the performance of the \textsc{OCaml}
  byte-code interpreter using Just-In-Time native code generation.
  Our JIT engine \textsc{OCamlJit2} currently runs
  on x86-64 processors, mimicing precisely the behavior of the
  \textsc{OCaml} virtual machine. Its design and implementation
  is described, and performance measures are given.
\end{abstract}

\section{Introduction}

The \textsc{OCaml} \cite{Leroy10,Remy02} system is the main implementation of the Caml
language \cite{Caml10}, featuring a powerful module system combined
with a full-fledged object-oriented layer. It comes with an optimizing native code compiler
\texttt{ocamlopt}, for high performance; a byte-code compiler \texttt{ocamlc}, for
increased portability; and an interactive top-level \texttt{ocaml}, for experimentation and rapid
development.

\texttt{ocamlc} and \texttt{ocaml} translate the source code into a sequence of byte-code 
instructions for the \textsc{OCaml} virtual machine, which is based on the ZINC machine
\cite{Leroy90} originally developed for Caml Light \cite{Leroy02}.
The optimizing native code compiler \texttt{ocamlopt} produces fast machine code for
the supported targets, but is only applicable to \emph{static program compilation}.
It cannot be used with multi-stage programming in \textsc{MetaOCaml} \cite{Taha06},
or the interactive top-level \texttt{ocaml}.

This paper introduces \textsc{OCamlJit2}, a new open-source\footnote{The full source code
is available from \url{https://github.com/bmeurer/ocamljit2/} under the terms of the Q Public
License and the GNU Library General Public License.} implementation of the \textsc{OCaml}
virtual machine, based on Just-In-Time native code generation. \textsc{OCamlJit2} is developed
on 64-bit Mac OS X 10.6, but should also run on Linux/amd64 and FreeBSD/amd64 systems without
modifications.

The paper is organized as follows: Section~\ref{section:Existing_systems} presents the
existing systems, including the relevant parts of the \textsc{OCaml} implementation, and the
previous \textsc{OCamlJit} implementation \cite{Starynkevitch04} which inspired the present work.
Section~\ref{section:Design_and_implementation} details the design and implementation
of \textsc{OCamlJit2}. Performance measures are given in section~\ref{section:Performance}.
Section~\ref{section:Related_work} and \ref{section:Future_work} list related and future
work, followed by the conclusion in section~\ref{section:Conclusion}.

\section{Existing systems} \label{section:Existing_systems}

We present the existing \textsc{OCaml} system and the previous \textsc{OCamlJit} \cite{Starynkevitch04}
implementation, which is based on the \textsc{Gnu Lightning} library \cite{Bonzini10}. For the sake of completeness
we repeat the overview of the \textsc{OCaml} compiler and runtime given in \cite{Starynkevitch04} below.
Readers already familiar with the internals of the \textsc{Ocaml} implementation can skip to
\ref{subsection:OCamlJit}.

\subsection{OCaml compiler usage}

The \textsc{OCaml} system can be used either interactively via the \texttt{ocaml} command, which
provides a ``\textit{read, compile, eval}'' loop, or in batch mode via the \texttt{ocamlc}
command, which takes a set of \texttt{*.ml} or \texttt{*.mli} source or \texttt{*.cmi} or
\texttt{*.cmo} byte-code object files\footnote{Source files are either module interfaces
  \texttt{*.mli} or module implementations \texttt{*.ml}. Byte-code object files are either
  compiled module interfaces \texttt{*.cmi} or compiled module implementations \texttt{*.cmo}.}
and produces a byte-code object file (to be reused by a subsequent invocation of \texttt{ocamlc})
or a byte-code executable file. \texttt{ocamlc} also handles \texttt{*.c} sources and \texttt{*.so}
shared libraries (for external C functions invoked from \textsc{OCaml}) and also deals with
byte-code library files \texttt{*.cma}. The byte-code executable files produced by \texttt{ocamlc}
are interpreted by the \texttt{ocamlrun} program\footnote{Technically speaking, an \textsc{OCaml}
  byte-code executable file is a \texttt{\#!/usr/local/bin/ocamlrun} Unix script, so \texttt{execve}
  of an \textsc{OCaml} byte-code executables invokes \texttt{ocamlrun}.} upon execution.

The goal of the present work is to substitute the interpreter part of the \textsc{OCaml} runtime
\texttt{ocamlrun} with a Just-In-Time compiler, which incrementally compiles the byte-code to
native code and runs the generated native code, instead of interpreting the byte-code. Unlike
\textsc{OCamlJit} \cite{Starynkevitch04}, we do not aim to provide a separate \texttt{ocamljitrun}
runtime, since that way one would have to explicitly use the runtime replacement to achieve the
benefits of Just-In-Time compilation.

\subsection{Overview of the OCaml system}

This section gives a brief overview of the \textsc{OCaml} byte-code compiler and runtime.
Feel free to skip to \ref{subsection:OCamlJit} if you are already familiar with the details.

\subsubsection{Compiler phases and representations}

\begin{figure}[ht]
  \centering
  \begin{tikzpicture}
    \node {$\mathbf{let}$}
    child {node {$\mathbf{var}\,x$}}
    child {node {$\mathbf{const}\,1$}}
    child {node {$\mathbf{apply}$}
      child {node {$\mathbf{var}\,+$}}
      child {node {$\mathbf{var}\,x$}}
      child {node {$\mathbf{const}\,3$}}
    };
  \end{tikzpicture}
  \caption{Abstract syntax tree}
  \label{figure:Abstract_syntax_tree}
\end{figure}
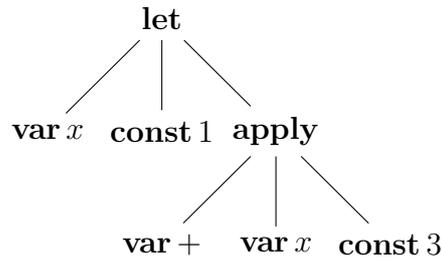

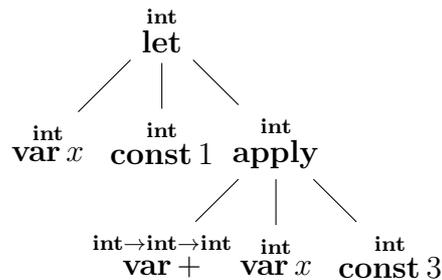
\begin{figure}[ht]
  \centering
  \begin{tikzpicture}
    \node {$\stackrel{\mathbf{int}}{\mathbf{let}}$}
    child {node {$\stackrel{\mathbf{int}}{\mathbf{var}\,x}$}}
    child {node {$\stackrel{\mathbf{int}}{\mathbf{const}\,1}$}}
    child {node {$\stackrel{\mathbf{int}}{\mathbf{apply}}$}
      child {node {$\stackrel{\mathbf{int}\to\mathbf{int}\to\mathbf{int}}{\mathbf{var}\,+}$}}
      child {node {$\stackrel{\mathbf{int}}{\mathbf{var}\,x}$}}
      child {node {$\stackrel{\mathbf{int}}{\mathbf{const}\,3}$}}
    };
  \end{tikzpicture}
  \caption{Typed syntax tree}
  \label{figure:Typed_syntax_tree}
\end{figure}

Compilation starts by parsing an \textsc{OCaml} source file (or a source region in interactive
mode) into an abstract syntax tree (AST, see file \texttt{parsing/parsedtree.mli} of the \textsc{OCaml}
source code). Figure~\ref{figure:Abstract_syntax_tree} illustrates the abstract syntax tree for the expression
\texttt{let x=1 in x+3}. Compilation then proceeds by computing the type annotations to produce
a typed syntax tree (see file \texttt{typing/typedtree.mli}) as shown in Figure~\ref{figure:Typed_syntax_tree}.

From this typed syntax tree, the byte-code compiler generates a so called {\em lambda representation} (see
file \texttt{bytecomp/lambda.mli})\footnote{The native code compiler \texttt{ocamlopt} uses the same
  intermediate lambda representation, but also includes an additional representation (see file
  \texttt{asmcomp/clambda.mli}), which adds explicit direct or indirect calls and closures.}
as shown in Figure~\ref{figure:Lambda_representation}, inspired by
the untyped call-by-value $\lambda$-calculus. This intermediate representation is not directly related to the
source code, because source file positions and some names are lost at this stage, and does not contain any
kind of explicit type information.

This lambda tree representation includes leaves for variables and constants as well as nodes for function
definitions and applications, \texttt{let} and \texttt{let rec}, primitive operations (like addition, 
multiplication, array access, etc.), switches (used in compiled forms of pattern matchings),
exception handling (\texttt{raise} and \texttt{try$\ldots$with} nodes), imperative traits (sequences,
\texttt{for} and \texttt{while} loops), etc.

\begin{figure}[ht]
  \centering
  \texttt{(let (x/1038 1) (+ x/1038 3))}
  \caption{Lambda representation}
  \label{figure:Lambda_representation}
\end{figure}

After several optimizations are applied (mostly peephole optimizations, transforming lambda
trees into \emph{better} or smaller lambda trees), the lambda tree representation is
transformed into a list of byte-code instructions as shown in Figure~\ref{figure:Byte_code}.
This instruction list is optimized and afterwards written into the generated byte-code file
(or kept in a memory buffer for the interactive top-level).

\begin{figure}[ht]
  \centering
  \begin{tabular}{l}
    \texttt{const 1} \\
    \texttt{push} \\
    \texttt{acc 0} \\
    \texttt{offsetint 3}
  \end{tabular}
  \caption{Byte-code}
  \label{figure:Byte_code}
\end{figure}

The \texttt{ocaml} top-level and the \texttt{ocamlc} compiler provide two undocumented command
line options \texttt{-dlambda} and \texttt{-dinstr} to display the internal representations
mentioned above.

\subsubsection{The OCaml virtual machine}

The \textsc{OCaml} virtual machine (see file \texttt{byterun/interp.c}) is an interpreter for the
byte-code\footnote{Since each token of the byte-code is a 32-bit word, the byte-code is actually
  a \emph{word-code}.} produced by \texttt{ocamlc} (as described in the previous section). It operates on a 
stack of \textsc{OCaml} values and five \emph{virtual registers}: the stack pointer \texttt{sp},
the accumulator \texttt{accu}, the environment pointer \texttt{env} (pointing to the current
closure, which contains the values for the free variables),
the extra arguments counter \texttt{extra\_args} (for partial applications), and the byte-code program
counter \texttt{pc}. The virtual machine also deals with byte-code segments (byte-code sequences to be interpreted)
and a global data array \texttt{caml\_global\_data} of \textsc{OCaml} values. There is usually
only a single byte-code segment, the byte-code sequence in the executable file produced by
\texttt{ocamlc}, which also contains the marshalled representation of the global data.

\textsc{OCaml} values are either pointers (usually pointing to blocks in the garbage collected
heap) or tagged integers. Each block starts with a header containing a tag, the size of the block
and the color (used by the garbage collector). Some tags describe special blocks like strings,
closures or floating point arrays, but most of them are used for representing sum types.
\textsc{OCaml} distinguishes pointers and tagged integers using the least significant bit: if
the least significant bit is $1$, the value is a tagged integer, with the integer value stored
in the remaining $31$ or $63$ bits, otherwise the value is a pointer.

The \textsc{OCaml} stack, which is -- in contrast to the optimizing native code compiler -- separate from
the native C stack, contains values, byte-code return addresses and extra argument counts,
organized into so-called \emph{call frames}. The byte-code is pointed to from either return
addresses stored on the stack or closures stored in the heap. The first slot of every closure
contains the byte-code address of the closure's function code; the remaining slots contain
the values of the free variables, thereby forming the environment of the closure. Mutually
recursive closures -- used to implement the \texttt{let rec} language construct -- may contain
more than just one byte-code address, followed by the values of the free variables.

The byte-code interpreter \texttt{ocamlrun} starts by unmarshalling the global data, loading
the byte-code sequence into memory and processing dynamically linked libraries containing
external C function primitives referenced from the byte-code file. Once everything is in
place, the function \texttt{caml\_interprete} (see file \texttt{byterun/interp.c}) is invoked
to interpret the initial byte-code segment (using threaded code \cite{Bell73,ErtlGregg03}),
starting from its first byte-code, with an empty stack, and a default environment and accumulator.

\subsubsection{The byte-code instruction set}

Every byte-code instruction is represented by one or more consecutive 32-bit words. The
first word of each byte-code includes the operation (see file \texttt{byterun/instruct.h})
and the remaining words include the operands, which are constant integer arguments (usually
offsets). Most byte-code instructions operate on the accumulator and the top-most stack elements and
produce a result in the accumulator. Some byte-codes include offsets (denoted by $p,q,\ldots$
subscripts), which are either encoded within the operation or following the
operation as 32-bit words. All byte-code offsets referencing other byte-codes are interpreted relative to their
respective positions, so the byte-code forms a \emph{position-independent code}.

Byte-code instructions are classified as follows:
\begin{itemize}
\item Stack manipulation instructions: $\mathtt{ACC}_p$ loads \texttt{accu} with the value of the $p$-th top-most stack
  cell; $\mathtt{PUSH}$ pushes the accumulator value onto the stack; $\mathtt{POP}_p$ pops the top-most $p$ elements
  off the stack; $\mathtt{ASSIGN}_p$ places the accumulator value into the $p$-th top-most stack cell.
\item Loading instructions: $\mathtt{CONSTINT}_p$ loads \texttt{accu} with the tagged integer $p$, whereas
  $\mathtt{ATOM}_p$ loads \texttt{accu} with the $p$-tagged atom\footnote{Atoms are pre-allocated $0$-sized
    blocks.}.
\item Primitive operations: unary $\mathtt{NEGINT}$ (negation) and $\mathtt{BOOLNOT}$ (logical negation)
  operate on \texttt{accu}; binary operations $\mathtt{ADDINT}$, $\mathtt{SUBINT}$, $\mathtt{MULINT}$,
  $\mathtt{DIVINT}$, $\mathtt{MODINT}$, $\mathtt{ANDINT}$, etc.~take the first operand in \texttt{accu}
  and the second one popped off the stack, and place the result into the accumulator, where division and
  modulus also check for $0$ and may raise an exception; comparisons $\mathtt{EQ}$, $\mathtt{NEQ}$,
  $\mathtt{LTINT}$, $\mathtt{LEINT}$, etc.~similarly place their boolean result into \texttt{accu};
  $\texttt{ISINT}$ tests whether the accumulator value is a tagged integer.
\item Environment operations: $\mathtt{ENVACC}_p$ loads \texttt{accu} with the $p$-th slot of
  the current environment \texttt{env}.
\item Apply operations are divided into two categories: $\mathtt{APPLY}_p$ creates a new call frame on
  the \textsc{OCaml} stack and jumps to the called code, whereas $\mathtt{APPTERM}_{p,q}$ ($q$ denotes the
  height of the current stack frame) performs a tail-call to the called code. In either case the accumulator
  contains the closure which is applied to the $p$ top-most arguments on the \textsc{OCaml} stack and becomes
  the new \texttt{env}. Call frames contain arguments, byte-code return address, previous \texttt{env}, and previous
  \texttt{extra\_args}; when adding call frames, the stack may grow if necessary.
\item Function return instructions: $\mathtt{RETURN}_p$ pops the top-most $p$ values off the stack and
  returns to the caller; $\mathtt{RESTART}$ and $\mathtt{GRAB}_p$ handle partial application (allocating
  appropriate closures as necessary).
\item Closure allocation instructions: $\mathtt{CLOSURE}_{p,q}$ allocates a single closure with $p$ variable slots and
  byte-code at offset $q$; $\mathtt{CLOSUREREC}_{p,q,k_1,\ldots,k_p}$ allocates a mutually recursive 
  closure with $p$ functions $k_1,\ldots,k_p$ and $q$ variable slots.
\item Allocation instructions: $\mathtt{MAKEBLOCK}_{p,q}$ creates a $p$-tagged block of $q$ values (first
  in \texttt{accu}, remaining popped off the stack); $\mathtt{MAKEFLOATBLOCK}_p$ creates an array of floats.
  The accumulator is loaded with the pointer to the newly allocated block.
  Every allocation may trigger a garbage collection.
\item Field access and modify instructions: $\mathtt{GETGLOBALFIELD}_{p,q}$ loads \texttt{accu} with the
  $q$-th field of the $p$-th global value; $\mathtt{GETFIELD}_p$ loads the accumulator with the $p$-th
  field of the block pointed to by \texttt{accu}; symmetrically $\mathtt{SETFIELD}_p$ sets the $p$-th
  field of \texttt{accu} to the value popped off the stack; similarly $\mathtt{GETFLOATFIELD}_p$ and
  $\mathtt{SETFLOATFIELD}_p$ handle fields in floating point blocks; $\mathtt{GETGLOBAL}_p$ and
  $\mathtt{SETGLOBAL}_p$ handle global fields in \texttt{caml\_global\_data}; $\mathtt{GETSTRINGCHAR}$
  and $\mathtt{SETSTRINGCHAR}$ are used to access and update characters within strings (the index
  being popped off the stack). All modifying operations have to cooperate with the garbage collector.
\item Control instructions include -- in addition to the instructions for function application --
  conditional $\mathtt{BRANCHIF}_p$ and $\mathtt{BRANCHIFNOT}_p$ (depending upon boolean value in
  \texttt{accu}), comparing $\mathtt{BEQ}_{p,q}$, $\mathtt{BNEG}_{p,q}, \ldots$ (comparing $q$ with
  the integer value in \texttt{accu}) and unconditional $\mathtt{BRANCH}_p$ jumps. There is also
  a $\mathtt{SWITCH}_{p,q,k_1,\ldots,k_p,k_1',\ldots,k_q'}$ instruction (used to compile the \textsc{OCaml}
  \texttt{match} construct), which tests \texttt{accu} and jumps to offset $k_{i-1}$ if \texttt{accu}
  is the tagged integer $i$ or $k_{j-1}'$ if \texttt{accu} points to a $j$-tagged block.
  Exception and signal handling instructions $\mathtt{PUSHTRAP}_p$, $\mathtt{POPTRAP}$, $\mathtt{RAISE}$
  $\mathtt{CHECK\_SIGNALS}$ are also control instructions, as is the halting $\mathtt{STOP}$ instruction,
  which is the last byte-code of a segment.
\item Calling external C primitives: $\mathtt{C\_CALL}_{p,q}$ calls the $p$-th C primitive with
  $q$ arguments (first taken from \texttt{accu}, remaining popped off the stack). The result
  is stored into \texttt{accu}. Primitives are used for several basic operations, including
  operations on floating point values.
\item Object-oriented operations: $\mathtt{GETMETHOD}$ retrieves an object method by its index;
  $\mathtt{GETPUBMET}$ and $\mathtt{GETDYNMET}$ fetch the method for a given method tag.
\item Debugger related instructions: the \textsc{OCaml} debugger places breakpoints by
  overwriting byte-code instructions with $\mathtt{BREAK}$; these are unsupported by
  \textsc{OCamlJit2}.
\end{itemize}

\subsection{OCamlJit} \label{subsection:OCamlJit}

In this section we briefly describe the design and implementation of the \textsc{Gnu Lightning} \cite{Bonzini10}
based \textsc{OCamlJit} \cite{Starynkevitch04},
which inspired many aspects of our present work. The main goal of \textsc{OCamlJit} is maximal compatibility
with the \textsc{OCaml} byte-code interpreter, its runtime system (including the gar\-bage collector), and its
behavior. That means programs running under \texttt{ocamljitrun} see the same virtual machine as
\texttt{ocamlrun}, so utilizing \textsc{OCamlJit} is a matter of exchanging \texttt{ocamlrun} with
\texttt{ocamljitrun} when executing \textsc{OCaml} programs.

In order to meet these constraint, \textsc{OCamlJit} has to mimic as much as possible the byte-code
interpreter.

\subsubsection{Implementation}

Since \textsc{Lightning} provides only a few registers, the most commonly used \textsc{OCaml} virtual
registers (\texttt{accu}, \texttt{sp}, \texttt{env}) are mapped to \textsc{Lightning} registers (hence
to machine registers), while other less common \textsc{OCaml} registers are grouped into a \emph{state
record} pointed to by a \textsc{Lightning} register.

The core of \textsc{OCamlJit} is the \texttt{caml\_jit\_translate} C function, which scans a byte-code
sequence and emits equivalent native code. It loops around a big \texttt{switch} statement, with one
\texttt{case} for every byte-code instruction. For example, Figure~\ref{figure:caml_jit_translate_andint} shows
the case for the \texttt{ANDINT} instruction, which pops a tagged integer off the \textsc{OCaml} stack
and does a ``bitwise and'' with \texttt{accu}.

\begin{figure}[ht]
  \centering
  \begin{varwidth}{\linewidth}
  \begin{verbatim}
case ANDINT:
  /* tmp1 = *sp; ++sp; accu &= tmp1; */
  jit_ldr_p(JML_REG_TMP1, JML_REG_SP);
  jit_addi_p(JML_REG_SP, JML_REG_SP, WORDSIZE);
  jit_andr_l(JML_REG_ACCU, JML_REG_ACCU,  JML_REG_TMP1);
  break;
\end{verbatim}
  \end{varwidth}
  \caption{\texttt{ANDINT} case in \texttt{caml\_jit\_translate}}
  \label{figure:caml_jit_translate_andint}
\end{figure}

\paragraph{Byte-code and native-code addresses}

Total compatibility demands the use of byte-code addresses, in particular within \textsc{OCaml}
closures. This means that closure application -- a very common operation in \textsc{OCaml} -- has
to retrieve the byte-code from the closure, find the corresponding native machine code, and jump to
it. The byte-code to native machine code address mapping is accomplished by a sparse 
translation hash-table. An entry is added to this table for every compiled byte-code instruction.

Every byte-code segment has its own native code block, which references a linked list of native
code chunks. Such a code chunk is a page-aligned executable memory segment, allocated via the
\texttt{mmap} system call with read, write and execute permissions.

The byte-code is incrementally translated to native machine code. The native code generation is
interrupted when the current native code chunk is filled, a \texttt{STOP} instruction is
reached, a configurable number of byte-code instructions has been translated, or when the
currently translated byte-code address is already known in the translation hash-table
(because a byte-code sequence containing it was previously translated).

\paragraph{Interaction with the runtime}

The generated native machine code has to interact with the runtime. A common interaction is
invoking the garbage collector when space in the minor heap is exhausted during allocations
(i.e.~\texttt{MAKEBLOCK}, \texttt{CLOSURE}, etc.). Most of the commonly used runtime interactions
are inlined in the generated native code.

Less common interactions, which are too complex to be inlined, are performed by saving the current
state of the virtual machine, returning from the virtual machine with a particular
\texttt{MLJSTATE\_*} (passing the required data in the state structure), and letting the
\texttt{caml\_interprete} function do the processing. For example \texttt{MLJSTATE\_MODIF} is
used to modify a heap cell while \texttt{MLJSTATE\_RAISE} is used to raise an exception (via C).

\paragraph{Performance}

On x86 machines, \texttt{ocamljitrun} typically gives significant
speedups with respect to \texttt{ocamlrun} of a factor above two. While this is already a quite
interesting achievement, there are however some shortcomings to be noted:
\begin{enumerate}
\item The compilation overhead is high, even with this naive compilation scheme. This is
  especially noticeable in short running programs. For example, building the \textsc{OCaml}
  standard library is nearly three times slower with \texttt{ocamljitrun} than with
  \texttt{ocamlrun}. There are various points that add to this slow translation speed; for
  example, adding every (byte-code, native-code) address pair to the translation
  hash-table takes a significant amount of the translation time -- actually more than a fourth
  of it \cite{Starynkevitch04}.
\item The limited register set provided by \textsc{Gnu Lightning} prevents efficient register
  usage for modern targets like x86-64, ARM or PowerPC, which offer more than just 8 general
  purpose registers. In addition, the \emph{portability at the lowest level} approach of
  \textsc{Gnu Lightning} prevents interesting target specific optimizations.
\item The naive compilation scheme (with peephole optimizations) is limited in efficiency. There
  is almost no way to achieve performance close to the optimizing native code compiler \texttt{ocamlopt}.
\end{enumerate}
The present work is part of a research project to overcome these problems. 

\subsection{Runtime code generators}

Several runtime code generators are available today, including but not limited to
\textsc{Gnu Lightning} \cite{Bonzini10}, \textsc{LLVM} \cite{Lattner02,Lattner04}
and \textsc{AsmJit} \cite{Kobalicek10}. These projects provide building blocks for
JIT engines using different levels of abstraction.

\textsc{Gnu Lightning}, which was recently ported to the x86-64 architecture\footnote{Also
known as \emph{AMD64 architecture}, but we prefer the vendor-neutral term.}, provides
users with a \emph{portability at the lowest level} approach. To gain reasonable
portability, one has to limit itself to the least common denominator w.r.t.~instruction
sets and registers available on the target platforms. This is a great concept for
rapid prototyping, but it is bound to fail in the long run.

\textsc{LLVM}\footnote{\url{http://llvm.org/}} provides an internal code representation,
which describes a program using an abstract RISC-like instruction set, and offers C/C++
APIs for code generation and Just-In-Time compilation (currently supporting x86,
x86-64 and PowerPC). But this comes at a cost: In a simple \textsc{OCamlJit2} prototype
based on \textsc{LLVM}, we have measured significant compilation overhead; short running
programs (like \texttt{ocamlc} applied to a small to medium sized \texttt{*.ml} file)
were three to four times slower than running the same program with the byte-code
interpreter. Similar results were observed by other projects using \textsc{LLVM} for
Just-In-Time compilation. For example, the Mono project\footnote{\url{http://www.mono-project.com/}}
reported impressive speedups for long running, computationally intensive applications,
but at the cost of increased compilation time and memory usage.

The \textsc{AsmJit} project aims to provide a nice API to emit x86 and x86-64 code
at runtime (support for ARM processors is planned) using a well-designed C++ API. Our
first fully working prototype used \textsc{AsmJit} to generate x86-64 native code.
While this worked out very well, we came to the conclusion that we did not actually
need any of the features provided by \textsc{AsmJit}, except for the code emitter
functions, and we found mixing C++ code with the somewhat ancient \textsc{OCaml} C
code a rather painful experience. We ended up writing our own C preprocessor macros
replacing the code emitter functions provided by \textsc{AsmJit}.

\section{Design and implementation} \label{section:Design_and_implementation}

We aim to provide almost the same amount of compatibility with the \textsc{OCaml} byte-code interpreter
as provided by \textsc{OCamlJit}, although not at all costs. At each point where we have to either
adjust the runtime or add costly work-arounds within the JIT engine, we choose to adjust the
runtime. Nonetheless \textsc{OCamlJit2} should be able to handle any byte-code sequence, and the
heap and stack should be the same as with the \textsc{OCaml} byte-code interpreter.

Our design goals are therefore similar to those of \textsc{OCamlJit}:
\begin{itemize}
\item The runtime (i.e.~the garbage collector and the required C primitives) should be 
  (roughly) the same as in the byte-code interpreter.
\item The heap is the same, in particular all \textsc{OCaml} values keep exactly the same representation
  (i.e.~the code pointer inside a closure is still a byte-code pointer, not a native code pointer). The
  relevant parts of the byte-code runtime (i.e.~the serialization mechanism) are reused\footnote{Especially
    hashing and serializing closures or objects (with their classes) give the same result
    in \textsc{OCamlJit2} as in the \textsc{OCaml} byte-code interpreter.}.
\item C programs embedding the \textsc{OCaml} byte-code interpreter should work with \linebreak[4]\textsc{OCamlJit2}
  as well, without the need to recompile the C program\footnote{We therefore need to ensure that the
    API of \textsc{OCamlJit2}'s \texttt{libcamlrun\_shared.so} match exactly the API of \textsc{OCaml}'s
    \texttt{libcamlrun\_shared.so}.}.
\item \textsc{OCamlJit2} should execute programs at least as fast as the byte-code interpreter,
  even for short running programs. In particular, the compilation time must not be noticeably
  longer when evaluating phrases within the \texttt{ocaml} top-level.
\end{itemize}
A program running with \textsc{OCamlJit2} should therefore see the same virtual machine as seen when
running with the byte-code interpreter. It should not be able to tell whether it is being interpreted
or JIT compiled, except by measuring the execution speed.

The relevant files are \texttt{byterun/jit.c}, which contains the actual JIT engine,
\linebreak[4]\texttt{byterun/jit\_rt\_amd64.S} containing the x86-64 runtime setup and helper routines
for the JIT generated machine code, and \texttt{byterun/jx86.h}, which contains the
x86-64 (and x86) code emitter preprocessor macros.

\subsection{Virtual machine design} \label{subsection:Virtual_machine_design}

This section describes the mapping of the \textsc{OCaml} virtual machine to the x86-64 architecture,
in particular the mapping of the virtual machine registers to the available physical registers. This
is inherently target specific, in contrast to most of the other ideas we present.

The x86-64 architecture \cite{Amd09Vol1,Intel10Vol1} provides $16$ general purpose 64-bit registers \texttt{\%rax},
\texttt{\%rbx}, \texttt{\%rcx}, \texttt{\%rdx}, \texttt{\%rbp}, \texttt{\%rsp}, \texttt{\%rdi}, \texttt{\%rsi},
\texttt{\%r8}, $\ldots$, \texttt{\%r15}, as well as $8$ 80-bit floating-point registers (also used
as MMX registers) and $16$ 128-bit SSE2 registers \texttt{\%xmm0}, $\ldots$, \texttt{\%xmm15}.
The System V ABI for the x86-64 architecture \cite{Matz10}, implemented by almost all operating
systems running on x86-64 processors\footnote{With the notable exception of
Win64, which uses a quite different ABI.}, mandates that registers \texttt{\%rbx}, \texttt{\%rbp},
\texttt{\%rsp} and \texttt{\%r12} through \texttt{\%r15} belong to the calling function and the called
function is required to preserve their values. The remaining registers belong to the called function.

Since the native machine code generated by the JIT engine will have to call C functions (i.e.~primitives
of the byte-code runtime) frequently, it makes sense to make good use of callee-save
registers to avoid saving and restoring too many aspects of the machine state whenever a C function
is called. We therefore settled on the following assignment for our current implementation:
\begin{itemize}
\item \texttt{accu} goes into \texttt{\%rax} for various reasons: typically \texttt{accu} does not
  need to be preserved across C calls, but is usually assigned the result of C function calls, which
  is then already available in \texttt{\%rax} \cite{Matz10}; additionally, many computations are performed on
  \texttt{accu} and several x86-64 instructions have shorter encodings when used with \texttt{\%rax}.
\item \texttt{extra\_args} goes into \texttt{\%r13}. One difference to the byte-code interpreter
  is that \texttt{\%r13} contains the number of extra arguments as tagged integer, in order to
  avoid the conversions when storing it to and loading it off the \textsc{OCaml} stack.
\item The environment pointer \texttt{env} goes into \texttt{\%r12}.
\item The stack pointer \texttt{sp} goes into \texttt{\%r14}.
\item The byte-code program counter \texttt{pc} is not needed anymore.
\end{itemize}
We use \texttt{\%r15} to cache the value of the minor heap allocation pointer \texttt{caml\_young\_ptr},
similar to \texttt{ocamlopt}. This greatly speeds up allocations of blocks in the minor heap,
which is a common operation. The remaining registers -- except for \texttt{\%rsp}, of course --
are available to implement the native machine code for the byte-code instructions.

There are several other target specific issues to resolve besides the assignment of registers. For example,
we need to ensure that the native stack is always aligned on a $16$-byte boundary prior to calling C functions.
But those issues are beyond the scope of this document.

The setup and helper routines that form the runtime of the JIT engine are in the file
\texttt{byterun/jit\_rt\_amd64.S}. The most important is \texttt{caml\_jit\_rt\_start}, which
starts the virtual machine by setting up the runtime environment, loading the appropriate registers,
and jumping to the first byte-code instruction (via a byte-code trampoline). This is the only
JIT runtime function that may be called from C, all other functions may only be
called from inside the generated native code. \texttt{caml\_interprete} invokes \texttt{caml\_jit\_rt\_start}
to execute the byte-code instruction sequence given to it (using a demand driven compilation
scheme, described below).

When the execution of the byte-code instruction sequence is done (i.e.~the translated \texttt{STOP}
instruction is reached) or an exception is thrown and not caught within the byte-code sequence,
\texttt{caml\_jit\_rt\_stop} is invoked, which undoes the runtime setup of \texttt{caml\_jit\_rt\_start}
and returns to the calling C function (i.e.~\texttt{caml\_interpret}).

\subsection{Address mapping} \label{subsection:Address_mapping}

As previously described, \textsc{OCamlJit} uses a sparse hash-table to map byte-code addresses to
native machine code addresses.
Whenever the virtual machine needs to jump to a byte-code pointer (i.e.~during closure application),
it has to lookup the corresponding native machine code in the hash-table, and jump to it (falling back
to the JIT compiler if no native code address is recorded).

Since closure application is a somewhat common operation in \textsc{OCaml}, this indirection via a
hash-table does not only complicate the implementation, but also decreases the performance of the
virtual machine. In addition, constructing the hash-table during JIT compilation takes a significant
amount of the translation time. To make matters worse, \textsc{OCamlJit} adds an entry for every
translated (byte-code, native-code) address pair to the hash-table, even though the vast majority
of these entries is never used.

Using a hash-table for the address mapping was therefore not an option for \textsc{OCamlJit2}. While
looking for another way to associate native code with byte-code addresses, we found that the threaded
byte-code interpreter is faced with a similar problem, whose solution may also be applicable in the
context of a JIT engine. When threading is enabled for the byte-code interpreter\footnote{This requires
the byte-code interpreter to be compiled with \texttt{gcc} at the time of this writing.}, prior to
execution, every byte-code segment is preprocessed by the C function \texttt{caml\_thread\_code},
which replaces each instruction opcode in the byte-code sequence with the offset of the C code
implementing the instruction relative to some well-defined base address. Executing a byte-code
instruction is then a matter of determining the C code address from the offset in the instruction
opcode and the well-defined base address\footnote{On 32-bit targets the base address is $0$, hence the
offset is already the address, while on 64-bit targets the offset is usually added to the base address.},
and jumping to that address. The threaded byte-code sequence itself therefore provides a mapping
from byte-code to native machine code addresses.

We have found that a similar approach is applicable in the context of a JIT engine. Whenever a
given byte-code instruction is JIT compiled to native code, we replace the instruction opcode word
within the byte-code sequence by the offset of the generated native code relative to the base address
\texttt{caml\_jit\_code\_end}. When jumping to a byte-code address -- i.e.~during closure application -- we
just read the offset from the instruction opcode field, add it to \texttt{caml\_jit\_code\_end},
and jump to that address. This trampoline code for x86-64 is shown in Figure~\ref{figure:Byte_code_trampoline_x86_64}
(\texttt{\%rdi} contains the address of the byte-code to execute and \texttt{\%rdx} is a temporary
register, not preserved across byte-code jumps).

\begin{figure}[ht]
  \centering
  \begin{varwidth}{\linewidth}
  \begin{verbatim}
movslq (%rdi), %rdx
addq   caml_jit_code_end(%rip), %rdx
jmpq   *%rdx
\end{verbatim}
  \end{varwidth}
  \caption{Byte code trampoline (x86-64)}
  \label{figure:Byte_code_trampoline_x86_64}
\end{figure}

Care must be taken on 64-bit targets, as the addressable range is limited to \mbox{$-2^{31} \ldots 2^{31}-1$} bytes,
since the width of the instruction opcode field is only 32 bits. Our current approach allocates
a fixed $2^{27}-x$ bytes code area\footnote{Where $x$ is a small amount of memory reserved for special purposes.}
(using \texttt{mmap}) with \texttt{caml\_jit\_code\_end} pointing to the end of the code area.
The offsets are therefore values in the range $-2^{27}+x$ to $-1$.

\subsection{Demand driven compilation}

Byte-code is incrementally translated to native machine code. Whenever the JIT engine detects a
byte-code instruction, which has not been translated to native machine code yet, it invokes the
C function \texttt{caml\_jit\_compile}. \texttt{caml\_jit\_compile} then translates the byte-code
block starting at the given byte-code offset and all byte-code blocks, which appear as branch
targets. These \emph{byte-code blocks} are basic blocks within the \textsc{OCaml} byte-code,
terminated by branch, raise, return and stop instructions. Translation of a byte-code block also
stops once an instruction is hit, which was already translated, and the compiler simply generates
a jump to the previously translated native machine code.

With the address mapping described above, this incremental, \emph{on-demand} compilation scheme
was very easy to implement. Given that all offsets to native machine code are negative values and
all byte-code instruction opcodes are positive values between $0$ and $145$, it is easy to distinguish
already translated and not yet translated byte-code instructions. Now one could alter the byte code
trampoline in Figure~\ref{figure:Byte_code_trampoline_x86_64} and add a test to see whether \texttt{(\%rdi)}
contains a positive value, and if so invoke the \texttt{caml\_jit\_compile} function.
We did however discover that such a test is unnecessary if we ensure that a jump to an address in
the range $0 \ldots 145$ relative to \texttt{caml\_jit\_code\_end} automatically invokes the
\texttt{caml\_jit\_compile} function. This was surprisingly easy to accomplish on x86-64 (and
x86); we simply reserve space after \texttt{caml\_jit\_code\_end} for $145$ \texttt{nop} instructions
and trampoline code which invokes \texttt{caml\_jit\_compile} and jumps to the generated code afterwards.
The byte-code compile trampoline for x86-64 is shown in Figure~\ref{figure:Byte_code_compile_trampoline_x86_64}.

\begin{figure}[ht]
  \centering
  \begin{varwidth}{\linewidth}
  \begin{verbatim}
movq  %rax, %rbx
call  caml_jit_compile(%rip)
xchgq %rbx, %rax
jmpq  *%rbx
\end{verbatim}
  \end{varwidth}
  \caption{Byte-code compile trampoline (x86-64)}
  \label{figure:Byte_code_compile_trampoline_x86_64}
\end{figure}

Now whenever a jump to a not yet translated byte-code instruction is performed, the byte-code trampoline
will jump to one of the \texttt{nop} instructions and fall through the following \texttt{nop}'s to the
byte-code compile trampoline. At this point \texttt{\%rax} contains the \texttt{accu} value, \texttt{\%rdi}
contains the byte-code address and current state of the virtual machine is stored in global variables
and callee-save registers (see section~\ref{subsection:Virtual_machine_design}).
Now the compile trampoline preserves \texttt{\%rax} in the
callee-save register \texttt{\%rbx} and invokes \texttt{caml\_jit\_compile} with the byte-code address to
compile in \texttt{\%rdi}. Once \texttt{caml\_jit\_compile} returns, \texttt{\%rax} contains the
address of the generated native code, so we save that address to \texttt{\%rbx}, restore the previous
value of \texttt{\%rax}, and jump to the native machine code.

\subsection{Main compile loop}

In the previous section we have already briefly described the \texttt{caml\_jit\_compile} function,
which performs the actual translation of byte-code instructions to native machine code. This section
gives a detailed overview of the translation process.

Whenever the byte-code compile trampoline invokes \texttt{caml\_jit\_compile}, it translates the byte-code
block starting at the byte-code instruction that triggered the translation (usually the beginning of a function,
i.e.~a \texttt{GRAB} instruction)
and continues until a terminating instruction is found (i.e.~\texttt{STOP}, \texttt{RETURN}, etc.) or an already
translated instruction is reached. During this process, all referenced byte-code addresses, which have no associated 
native machine code yet, are collected in a list of pending byte-code blocks. Once the translation of a byte-code
block is done, \texttt{caml\_jit\_compile} continues with the next pending block, and repeats until
all pending blocks are translated. It is trivial to see that this process terminates, since 
every block is translated at most once; as soon as it is translated, the native machine code address
is known and it will not be added to the pending list again.

The actual translation of byte-code instructions is done in a large \texttt{switch} statement, which is
quite similar to the non-threaded byte-code interpreter, except that we do not interpret the opcodes,
but generate appropriate native machine code. Figure~\ref{figure:caml_jit_compile_ANDINT} shows the 
case for the \texttt{ANDINT} instruction on x86-64.

\begin{figure}[ht]
  \centering
  \begin{varwidth}{\linewidth}
  \begin{verbatim}
case ANDINT:
  jx86_andq_reg_membase(cp, JX86_RAX, JX86_R14, 0);
  jx86_addq_reg_imm(cp, JX86_R14, 8);
  break;
\end{verbatim}
  \end{varwidth}
  \caption{\texttt{caml\_jit\_compile} \texttt{ANDINT} case (x86-64)}
  \label{figure:caml_jit_compile_ANDINT}
\end{figure}

The generated native code for \texttt{ANDINT} on x86-64 is shown in Figure~\ref{figure:ANDINT_native_code}.
The \texttt{andq} instruction performs the ``bitwise and'' of \texttt{accu} (located in \texttt{\%rax}) and
the top-most stack cell (located in \texttt{0(\%r14)}), storing the result into \texttt{\%rax}. The \texttt{addq}
instruction pops the top-most cell off the stack (the \textsc{OCaml} stack grows towards lower addresses).

\begin{figure}[ht]
  \centering
  \begin{varwidth}{\linewidth}
  \begin{verbatim}
andq 0(%r14), %rax
addq $8, %r14
\end{verbatim}
  \end{varwidth}
  \caption{\texttt{ANDINT} native code (x86-64)}
  \label{figure:ANDINT_native_code}
\end{figure}

The translation of most other byte-code instructions is just as straight-forward as
the \texttt{ANDINT} case shown above. There are however a few byte-code instructions that
required special processing, most notably branch instructions. These instructions are used
to change the flow of control within a byte-code segment.
Whenever a \texttt{BRANCH} instruction is hit, we check whether the branch target is already
translated (i.e.~whether the opcode of the instruction contains a negative code offset). If so, we
simply generate a \texttt{jmp} to the known native code address.
Otherwise we generate a \emph{forward jump} (a \texttt{jmp} instruction with no known
target), record a reference from the target byte-code to the forward jump\footnote{The
term \emph{forward jump} may be misleading, since the jump does not need to be forward
within the byte-code segment.}, and append the target byte-code to the pending list
(if it wasn't already listed there). The translation of conditional branch and \texttt{SWITCH}
instructions follows a similar scheme. Since \texttt{caml\_jit\_compile}
loops until all pending blocks are translated, we are sure that all forward jumps will
have been patched once \texttt{caml\_jit\_compile} returns.

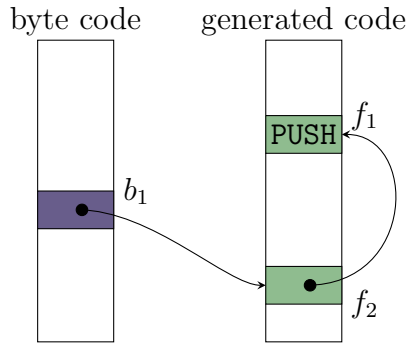
\begin{figure}[htb]
  \centering
  \definecolor{bytecolor}{HTML}{685D8B}
  \definecolor{nativecolor}{HTML}{8FBC8F}
  \begin{tikzpicture}
    \draw (0.5,5.25) node {byte code};
    \draw (3.5,5.25) node {generated code};
    \draw (0,1) rectangle (1,5);
    \filldraw [fill=bytecolor] (0,2.5) rectangle (1,3);
    \draw (3,1) rectangle (4,5);
    \filldraw [fill=nativecolor] (3,1.5) rectangle (4,2);
    \filldraw [fill=nativecolor] (3,3.5) rectangle (4,4);
    \draw [*-stealth] (0.5,2.75) .. controls (1.5,2.75) and (2.5,1.75) .. (3,1.75);
    \draw [*-stealth] (3.5,1.75) .. controls (5,1.75) and (5,3.75) .. (4,3.75);
    \draw (1.3,3) node {$b_1$};
    \draw (4.3,4) node {$f_1$};
    \draw (4.3,1.5) node {$f_2$};
    \draw (3.5,3.75) node {$\mathtt{PUSH}$};
  \end{tikzpicture}
  \caption{Threading of forward jumps}
  \label{figure:Threading_of_forward_jumps}
\end{figure}

Recording a reference from a target byte-code to a forward jump is done by threading the
jump templates to the target byte-code: The current value of the target byte-code opcode
is stored into the forward jump and the target byte-code opcode field is set to point
to the address of the forward jump. This is illustrated in Figure~\ref{figure:Threading_of_forward_jumps},
which shows two forward jumps $f_1$ and $f_2$ threaded to a byte-code address $b_1$,
which contains a not yet translated \texttt{PUSH} instruction.

The \texttt{caml\_jit\_compile} driver loop works by allocating a native code address,
patching all forward jumps threaded to the current byte-code address\footnote{We intentionally
omitted the implementation details of threading and patching, and refer to the interested
reader to the source code in the file \texttt{byterun/jit.c}.}, replacing the
instruction opcode of the current byte-code with the offset to the native machine code,
and generating the native code for the byte-code instruction at the allocated address.
This way all byte-code instructions within the translated blocks contain the offsets
of the appropriate native machine code once \texttt{caml\_jit\_compile} returns.

\subsection{Optimizations}

We have applied several optimizations to improve the performance of the generated
native machine code. In this section we describe two important optimizations that
led to significant performance improvements.

\subsubsection{Stack pointer updates}

The compilation scheme described above is similar to the one implemented by \textsc{OCamlJit}
in that it generates a lot of unnecessary stack pointer updates. For every instruction
that pushes values to or pops values off the stack an appropriate update of the stack
pointer is generated.
For example, the byte-code sequence \texttt{PUSH ACC4 ADDINT} generates two stack pointer
updates, one decrement by \texttt{PUSH} and one increment by \texttt{ADDINT}, which are
unnecessary as the stack pointer will point to the initial stack cell again after executing
the three instructions.

In order to avoid these unnecessary updates, we introduced a \emph{stack offset} into
the compilation loop, which identifies the current top of stack relative the stack
pointer in the hardware register. Whenever the real stack pointer must reside in the
hardware register (i.e.~at the end of a byte-code block), we emit an instruction to
add the stack offset to the stack pointer, and reset the stack offset to $0$.

This required changes to the address handling in the compilation loop:
Byte-code addresses can only be mapped to native code addresses if the stack
offset is $0$ at this point in code generation, and when patching forward jumps,
the stack offset must be flushed first. Now not every translated byte-code
instruction gets mapped to a native code address, which may lead to duplicate
translation of the same byte-code instruction. However this duplicate
translation happens seldomly and affects mostly backward branches, which are
usually loop branches jumping to a \texttt{CHECK\_SIGNALS} byte-code instruction.
So we can almost reduce the probability of duplicate translation to zero by
ensuring that the stack offset is flushed prior to generating code for \texttt{CHECK\_SIGNALS}
instructions.

This way we were able to avoid nearly all unnecessary stack pointer updates, which
also reduced the size of the generated native machine code by $4-17\,\%$ in
our tests.

\subsubsection{Floating point operations}

Most floating point operations are implemented as C primitives in the byte-code
interpreter. Calling C primitives from the generated native code comes at a
certain cost -- i.e.~making the stack and minor allocation pointer available
to C code, and reloading it once the C function returns -- which significantly
decreases the performance of programs using a lot of floating point operations.
We have therefore decided to inline various floating point primitives (\texttt{caml\_add\_float},
\texttt{caml\_sub\_float}, etc.), which speed up certain floating point benchmarks by a factor of $1.7$.

\section{Performance} \label{section:Performance}

The performance measurements presented here are done on a MacBook 
with an Intel Core 2 Duo 2.4 GHz CPU (3 MiB L2 Cache), and 4 GiB 1067 MHz DDR3 RAM,
running Mac OS X 10.6.4. The C compiler is \texttt{gcc-4.2.1} (Apple Inc.
build 5664). The \textsc{OCaml} compiler is 3.12.0. The \textsc{OCamlJit2}
code is the tagged revision \texttt{ocamljit2-2010-tr1},
compiled with \texttt{gcc} optimization level \texttt{-O3}.

\begin{table*}[ht]
  \footnotesize
  \centering
  \begin{tabular}{l|rrr|rrr|l}
    \multicolumn{1}{l|}{\large invocation}
    & \multicolumn{3}{c|}{{\large time} (cpu sec.)}
    & \multicolumn{3}{c|}{\large speedup}
    & \multicolumn{1}{l}{\large notes} \\
    command
    & $t_{byt}$
    & $t_{jit}$
    & $t_{opt}$
    & $\sigma^{jit}_{byt}$
    & $\sigma^{opt}_{jit}$
    & $\sigma^{opt}_{byt}$
    \\
    \hline
\texttt{almabench} & $27.83$ & $10.05$ & $4.47$ & $2.77$ & $2.25$ & $6.22$ & number crunching \\
\texttt{almabench.unsafe} & $27.74$ & $9.76$ & $4.35$ & $2.84$ & $2.24$ & $6.38$ & number crunching {\tiny(no bounds check)} \\
\texttt{bdd} & $8.50$ & $1.93$ & $0.68$ & $4.41$ & $2.86$ & $12.60$ & binary decision diagram \\
\texttt{boyer} & $4.34$ & $1.68$ & $1.06$ & $2.59$ & $1.59$ & $4.11$ & term processing \\
\texttt{fft} & $5.72$ & $2.17$ & $0.64$ & $2.63$ & $3.39$ & $8.94$ & fast fourier transformation \\
\texttt{nucleic} & $14.78$ & $4.10$ & $0.80$ & $3.61$ & $5.13$ & $18.52$ & floating point \\
\texttt{quicksort} & $6.78$ & $1.27$ & $0.24$ & $5.34$ & $5.34$ & $28.48$ & array quicksort \\
\texttt{quicksort.unsafe} & $4.07$ & $0.93$ & $0.20$ & $4.36$ & $4.79$ & $20.86$ & array quicksort {\tiny(no bounds check)} \\
\texttt{soli} & $0.17$ & $0.03$ & $0.01$ & $5.09$ & $3.40$ & $17.30$ & tiny solitaire \\
\texttt{soli.unsafe} & $0.14$ & $0.02$ & $0.01$ & $6.27$ & $2.75$ & $17.25$ & tiny solitaire {\tiny(no bounds check)} \\
\texttt{sorts} & $19.29$ & $7.13$ & $3.71$ & $2.71$ & $1.92$ & $5.19$ & various sorting algorithms \\
\hline
\texttt{ocamlc -help} & $0.01$ & $0.01$ & $0.00$ & $1.14$ & $7.00$ & $8.00$ & {\tiny(short execution)} \\
\texttt{ocamlc -c format.ml} & $0.17$ & $0.08$ & $0.05$ & $2.16$ & $1.65$ & $3.57$ &  \\
\texttt{ocamlopt -c format.ml} & $0.65$ & $0.28$ & $0.19$ & $2.31$ & $1.46$ & $3.38$ &  \\
\texttt{ocamlc -c stdlib/*.ml} & $2.03$ & $0.95$ & $0.67$ & $2.14$ & $1.42$ & $3.04$ & byte-compile \texttt{stdlib/*.ml} \\
\texttt{ocamlopt -c stdlib/*.ml} & $5.04$ & $2.36$ & $1.70$ & $2.14$ & $1.39$ & $2.97$ & native-compile \texttt{stdlib/*.ml} \\
  \end{tabular}
  \caption{Running time and speedup}
  \label{table:Running_time_and_speedup}
\end{table*}

\subsection{Execution time speedup}

We measured only the total execution time, as this is the significant
time. Table~\ref{table:Running_time_and_speedup} gives some timings
(combined system and user CPU time, in seconds), comparing the byte-code interpreter
time $t_{byt}$ to the \textsc{OCamlJit2} time $t_{jit}$, and the
time $t_{opt}$ taken by the same program compiled with the optimizing
native code compiler \texttt{ocamlopt}. It also lists the relative speedups
$\sigma^{jit}_{byt} = \frac{t_{byt}}{t_{jit}}$, $\sigma^{opt}_{jit} = \frac{t_{jit}}{t_{opt}}$,
and $\sigma^{opt}_{byt} = \frac{t_{byt}}{t_{opt}}$, where bigger values are
better, and a value less than $1$ would represent a regression.

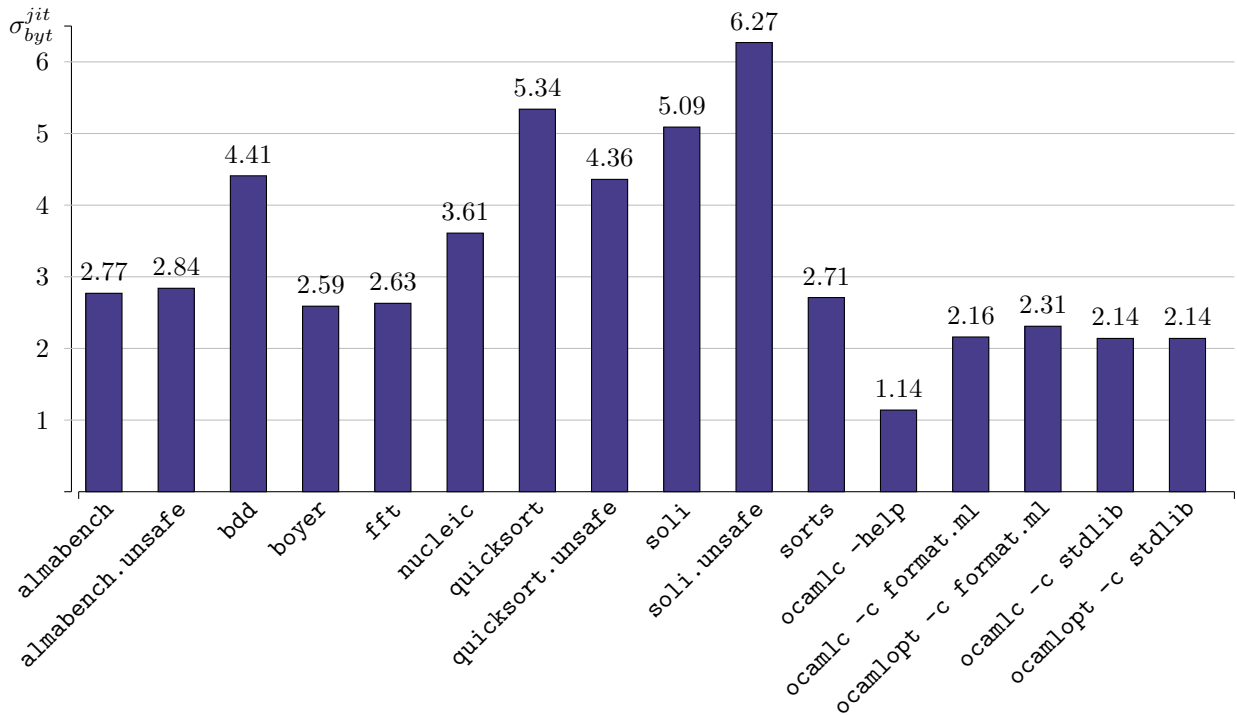
\begin{figure*}[bht]
  \centering
  \footnotesize
  \begin{tikzpicture}[scale=.95]
    \definecolor{color}{HTML}{483D8B}
    \draw (0cm,0cm) -- (16.0cm,0cm);  
    \draw (0cm,0cm) -- (0cm,-0.1cm);  
    \draw (16.0cm,0cm) -- (16.0cm,-0.1cm);  
    \draw (-0.1cm,0cm) -- (-0.1cm,6.5cm);  
    \draw (-0.1cm,0cm) -- (-0.2cm,0cm);  
    \draw (-0.1cm,6.5cm) -- (-0.2cm,6.5cm) node [left] {$\sigma^{jit}_{byt}$};  
    \foreach \x in {1,...,6}  
      \draw[gray!50, text=black] (-0.2 cm,\x cm) -- (16.0 cm,\x cm) node at (-0.5 cm,\x cm) {\x};  
    \foreach \x/\y/\bench in
      {
        0.1/2.77/almabench,  
        1.1/2.84/almabench.unsafe,  
        2.1/4.41/bdd,
        3.1/2.59/boyer,
        4.1/2.63/fft,
        5.1/3.61/nucleic,
        6.1/5.34/quicksort,
        7.1/4.36/quicksort.unsafe,
        8.1/5.09/soli,
        9.1/6.27/soli.unsafe,
        10.1/2.71/sorts,
        11.1/1.14/ocamlc -help,
        12.1/2.16/ocamlc -c format.ml,
        13.1/2.31/ocamlopt -c format.ml,
        14.1/2.14/ocamlc -c stdlib/*.ml,
        15.1/2.14/ocamlopt -c stdlib/*.ml
      }
      {
        \draw[fill=color] (\x cm,0cm) rectangle (0.5cm+\x cm,\y cm) 
        node at (0.25cm + \x cm,\y cm + 0.3cm) {\y}; 
        \node[rotate=45, left] at (0.4 cm +\x cm,-0.1cm) {\texttt{\bench}}; 
      };
  \end{tikzpicture}
  \caption{Speedup w.r.t.~the byte-code interpreter}
  \label{figure:Speedup_w_r_t_the_byte_code_interpreter}
\end{figure*}

The commands listed in the first $11$ rows are test programs from the
\texttt{testsuite/tests} folder of the \textsc{OCaml} 3.12.0 distribution.
They do more or less represent typical \textsc{OCaml} applications.
The remaining rows are invocations of the \texttt{ocamlc} and \texttt{ocamlopt}
compilers, where \texttt{stdlib/*.ml} means all source files from the
\texttt{stdlib} folder of the \textsc{OCaml} 3.12.0 distribution that
do not require special compilation arguments, like \texttt{-nopervasives}
for \texttt{stdlib/pervasives.ml*} or \texttt{-nolabels} for \texttt{stdlib/*Labels.ml*}.

Figure~\ref{figure:Speedup_w_r_t_the_byte_code_interpreter} highlights the
speedup of \textsc{OCamlJit2} w.r.t.~the byte-code interpreter. As expected,
really short executions (i.e.~\texttt{ocamlc -help}) do not benefit from the Just-In-Time
compilation, but thanks to our design, their running time is on par with the
byte-code interpreter (in contrast to \textsc{OCamlJit}, where the running
time was worse). Other invocations are speed up by a factor of $2.1$ to $6.3$.

Floating point intensive programs like \texttt{almabench}, \texttt{fft}
and \texttt{nucleic} spend a large amount of time collecting garbage,
which is due to the fact that we still allocate a heap block for each
and every floating point value used during execution. We intend to
further optimize the performance here using better instruction selection
and register allocation.

In particular short running programs, like invocations of \texttt{ocamlc}
with small to medium sized \texttt{*.ml} files, which caused noticeable
slow-downs with \textsc{OCamlJit}, benefit from \linebreak[4]\textsc{OCamlJit2}.
For example, building the full \textsc{OCaml}
distribution -- using \texttt{make world opt} in a clean source
tree -- takes only $153.8$s\footnote{Measured in combined user and
system CPU time.} with \textsc{OCamlJit2}, while
\textsc{OCaml} requires $261$s. This may not seem
significant upon first sight, but it is indeed a nice improvement
for a task that is mostly I/O bound and involves numerous invocations
of the C compiler, the assembler and the linker, which aren't
affected by the use of \textsc{OCamlJit2}.

\subsection{Branch prediction accuracy}

\begin{figure*}[bht]
  \centering
  \footnotesize
  \begin{tikzpicture}
    \definecolor{color1}{HTML}{8B0000}
    \definecolor{color2}{HTML}{4682B4}
    \draw (0cm,0cm) -- (8.5cm,0cm);  
    \draw (0cm,0cm) -- (0cm,-0.1cm);  
    \draw (8.5cm,0cm) -- (8.5cm,-0.1cm);  
    \draw (-0.1cm,0cm) -- (-0.1cm,4.5cm);  
    \draw (-0.1cm,0cm) -- (-0.2cm,0cm);  
    \draw (-0.1cm,4.5cm) -- (-0.2cm,4.5cm) node [left] {};  
    \foreach \x in {85,90,95,100}  
      \draw[gray!50, text=black] (-0.2 cm,\x * 0.2 cm - 80 * 0.2cm) -- (8.5 cm,\x * 0.2 cm - 80 * 0.2 cm) node at (-0.5 cm,\x * 0.2 cm - 80 * 0.2 cm) {$\x \%\quad$};  
    \foreach \x/\j/\b/\bench in
    {
      0/99.1/85.0/almabench,
      1/96.2/84.7/bdd,
      2/97.4/88.9/boyer,
      3/99.6/88.1/fft,
      4/98.9/89.2/nucleic,
      5/99.3/90.2/quicksort,
      6/98.0/89.6/soli,
      7/98.6/92.7/sorts
    }
    {
      \draw[fill=color1] (\x cm+0.1cm,0cm) rectangle (0.5cm+\x cm,\j * 0.2cm - 80 * 0.2cm);
      \draw[fill=color2] (\x cm+0.51cm,0cm) rectangle (0.9cm+\x cm,\b * 0.2cm - 80 * 0.2cm);
      \node[rotate=45,left] at (0.5cm + \x cm,-0.1cm) {\texttt{\bench}};
    };
    \draw[fill=color1] (9.5cm,4.0cm) rectangle (10.0cm,4.2cm)
    node at (11.1cm,4.1cm) {\textsc{OCamlJit2}};
    \draw[fill=color2] (9.5cm,3.6cm) rectangle (10.0cm,3.8cm)
    node at (10.745cm,3.7cm) {\textsc{OCaml}};
  \end{tikzpicture}
  \caption{Branch prediction accuracy}
  \label{figure:Branch_prediction_accuracy}
\end{figure*}
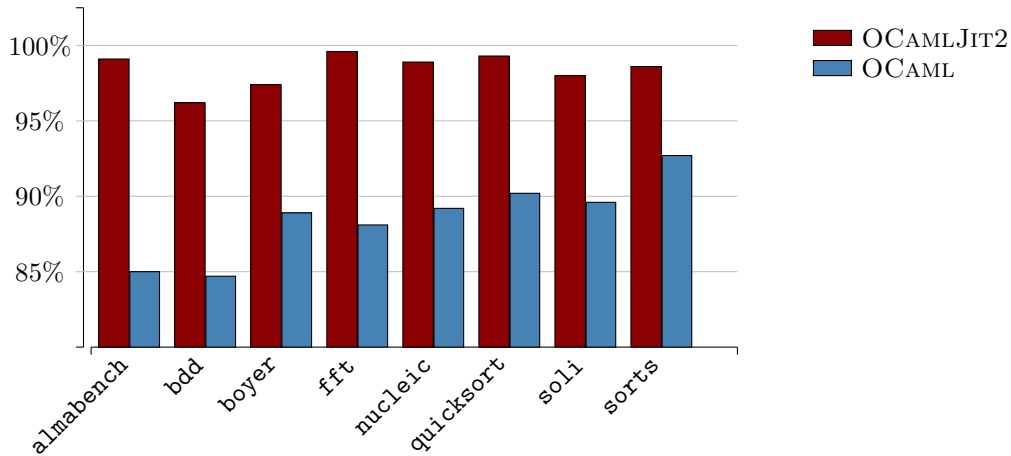

Efficient virtual machine interpreters -- like the \textsc{OCaml} byte-code interpreter --
perform a large number of indirect jumps (up to $11\%$ of all executed instructions in
our benchmarks). Even though branch target buffers (BTBs) in modern CPUs and various interpreter techniques
\cite{CaseyErtlGregg07} can greatly reduce the number of mispredicted branches, low branch prediction accuracy
is still a major bottleneck in virtual machine interpreters \cite{ErtlGregg01}.

Virtual machines based on Just-In-Time compilers are able to reduce almost all indirect branches and thereby
increase the branch prediction accuracy to nearly $98-99\%$ in most cases. A few indirect branches
cannot be eliminated, for example closure application in a virtual machine for a
functional language needs an indirect branch (unless the closure's code pointer is
known at compile time).
Figure~\ref{figure:Branch_prediction_accuracy} compares the branch
prediction accuracy of \textsc{OCaml} and \textsc{OCamlJit2} in various
benchmarks on the Intel Core 2 Duo.

\subsection{Comparison with OCamlJit}

Unfortunately we were unable to get \textsc{OCamlJit} running with \textsc{OCaml}
3.12.0 and the most recent revision of \textsc{Gnu Lightning}, which supports
x86-64 processors, so we are unable to directly compare the running times of
\textsc{OCamlJit} and \textsc{OCamlJit2} on identical hardware.

We can only compare the relative speedups achieved on different architectures
(x86 in case of \textsc{OCamlJit} and x86-64 in case of \textsc{OCamlJit2}),
which seem to indicate that we are doing quite well.

\section{Related work} \label{section:Related_work}

Just-In-Time compilation is an active research field \cite{Arnold04,Aycock03}.
A lot of work has been done on efficient dynamic compilation in the field of
the Java Virtual Machine (JVM) and the Common Language Runtime (CLR), which is
part of the .NET framework. JIT compilation of Java or .NET byte-code is usually
driven by first interpreting the byte-code, collecting profiling information,
and selecting the appropriate methods to optimize by inspecting the collected
profiling data \cite{Cuthbertson09,Vaswani03}.

Both JVM and CLR depend on profile guided optimizations for good performance,
because their byte-code is slower than the \textsc{OCaml} byte-code. This
is mostly because of the additional overhead that comes with exception handling,
multi-threading, class loading and object synchronization. But it is also due to
the fact that both the JVM byte-code \cite{LindholmYellin99} and the Common Intermediate
Language (CIL) \cite{Ecma335} instruction sets are not as optimized as the \textsc{OCaml} byte-code.

More recent examples include the Lua programming language \cite{Ierusalimschy07,Ierusalimschy96},
a powerful, fast, lightweight, embeddable scripting language, accompanied
by LuaJIT \cite{Pall10}, which combines a high-performance interpreter with
a state-of-the-art Just-In-Time compiler.

Another recent example is the Dalvik virtual machine\footnote{\url{http://dalvikvm.com/}}
\cite{Bornstein08}, which drives applications running on the Android mobile
operating system\footnote{\url{http://www.android.com/}}. As of Android 2.2,
Dalvik features a Just-In-Time compiler, which increased performance of
several applications by a factor of $2$ to $4$.

During the last decades a lot of work has been done on improving the
performance of interpreters and native code compilers for other
stack-based languages like \textsc{Forth} \cite{Ertl93}, in particular
compiling \textsc{Forth} to efficient native code for register
machines \cite{Ertl92,Ertl96,ErtlPirker97,ErtlPirker98}.

\section{Future work} \label{section:Future_work}

Although the current implementation already performs quite well in most cases,
it also has several drawbacks, which will be addressed in the future.

First of all, instruction selection, instruction scheduling and register
allocation \cite{Aho06} in the current implementation are far from optimal.
For example most of the generated code uses less than a fourth of the
registers provided by the x86-64 processor. This is due to the simple
compilation scheme. 

The code memory management is another open issue. Right now code memory
is managed by allocating in one large memory chunk. Unused code memory
is never released or reused, which may cause trouble for \textsc{MetaOCaml}
\cite{Taha06} or long-lived interactive top-level sessions. This could be
addressed by adding support for garbage collected byte-code segments to the
\textsc{OCaml} runtime, which would then also take care of native machine
code segments.

We also plan to improve the portability of the JIT engine. While it is more or
less straight-forward to port the current code to the x86 architecture (trivial
in most cases, but nevertheless a time consuming task), it would take considerable
amounts of time to port the JIT engine to a different architecture like PowerPC,
Sparc or ARM.

In the future, we may even be able to get the JIT engine on par with
the optimizing native code compiler \texttt{ocamlopt} w.r.t.~execution speed,
which would offer the possibility to drop \texttt{ocamlopt} and its runtime.
This would improve the maintainability of \textsc{OCaml}, because then
only a single compiler and a single runtime would have to be maintained
-- plus the JIT engine, of course. However this is an optional goal and
by no means a requirement, especially since \textsc{OCaml}'s optimizing
native code compiler is also of educational interest.

\section{Conclusion} \label{section:Conclusion}

The \textsc{OCamlJit2} implementation presented here did achieve some significant
speedups (at least twice as fast as the byte-code interpreter in all relevant
comparisons), but there are various open issues to address. In particular the
suboptimal instruction selection, instruction scheduling and register allocation
will have to be addressed to get \textsc{OCamlJit2} on par with the optimizing
native code compiler and JIT engines available for other programming languages
and runtime environments.

\section*{Acknowledgements}

We would like to thank Simon Meurer and Christian Uhrhan for their careful proof-reading.

\bibliographystyle{abbrv}
\bibliography{citations}

\begin{thebibliography}{10}

\bibitem{Amd09Vol1}
Advanced Micro Devices, Inc.
\newblock {\em AMD64 Architecture Programmer's Manual Volume 1: Application
  Programming}, Nov 2009.
\newblock
  \url{http://developer.amd.com/documentation/guides/Pages/default.aspx}.

\bibitem{Aho06}
A.~V. Aho, M.~S. Lam, R.~Sethi, and J.~D. Ullman.
\newblock {\em {C}ompilers: {P}rinciples, {T}echniques, and {T}ools}.
\newblock Addison Wesley, 2nd edition, August 2006.

\bibitem{Arnold04}
M.~Arnold, S.~J. Fink, D.~Grove, M.~Hind, and P.~F. Sweeney.
\newblock A {S}urvey of {A}daptive {O}ptimization in {V}irtual {M}achines.
\newblock In {\em Proceedings of the IEEE, 93(2), 2005. Special issue on
  program generation, optimization, and adaptation}, 2004.

\bibitem{Aycock03}
J.~Aycock.
\newblock A {B}rief {H}istory of {J}ust-{I}n-{T}ime.
\newblock {\em ACM Computing Surveys}, 35(2):97--113, 2003.

\bibitem{Bell73}
J.~R. Bell.
\newblock Threaded code.
\newblock {\em Communications of the {ACM}}, 16(6):370--372, 1973.

\bibitem{Bonzini10}
P.~Bonzini et~al.
\newblock {GNU} {L}ightning library.
\newblock \url{http://www.gnu.org/software/lightning/}, 2010.

\bibitem{Bornstein08}
D.~Bornstein.
\newblock Dalvik {VM} {I}nternals, May 2008.
\newblock \url{http://sites.google.com/site/io/dalvik-vm-internals/}.

\bibitem{Caml10}
The {C}aml {C}onsortium.
\newblock {\em The {C}aml {L}anguage}, 2010.
\newblock \url{http://caml.inria.fr/}.

\bibitem{CaseyErtlGregg07}
K.~Casey, M.~A. Ertl, and D.~Gregg.
\newblock Optimizing indirect branch prediction accuracy in virtual machine
  interpreters.
\newblock {\em ACM Trans. Program. Lang. Syst.}, 29, October 2007.

\bibitem{Cuthbertson09}
J.~Cuthbertson, S.~Viswanathan, K.~Bobrovsky, A.~Astapchuk, and E.~K.~U.
  Srinivasan.
\newblock A {P}ractical {A}pproach to {H}ardware {P}erformance {M}onitoring
  {B}ased {D}ynamic {O}ptimizations in a {P}roduction {JVM}.
\newblock In {\em CGO '09: Proceedings of the 7th annual IEEE/ACM International
  Symposium on Code Generation and Optimization}, pages 190--199, Washington,
  DC, USA, 2009. IEEE Computer Society.

\bibitem{Ecma335}
{ECMA} {I}nternational.
\newblock {\em {S}tandard {ECMA-335} - {C}ommon {L}anguage {I}nfrastructure
  {(CLI)}}, 4th edition, June 2006.
\newblock
  \url{http://www.ecma-international.org/publications/standards/Ecma-335.htm}.

\bibitem{Ertl92}
M.~A. Ertl.
\newblock A {N}ew {A}pproach to {Forth} {N}ative {C}ode {G}eneration.
\newblock In {\em EuroForth '92 Conference Proceedings}, pages 73--78,
  Southampton, England, 1992.

\bibitem{Ertl93}
M.~A. Ertl.
\newblock A portable {Forth} engine.
\newblock In {\em EuroForth '93 Conference Proceedings}, Mari\'ansk\'e
  L\'azn\`e (Marienbad), 1993.

\bibitem{Ertl96}
M.~A. Ertl.
\newblock {\em Implementation of Stack-Based Languages on Register Machines}.
\newblock PhD thesis, {Technische Universit\"{a}t Wien}, Austria, 1996.

\bibitem{ErtlGregg01}
M.~A. Ertl and D.~Gregg.
\newblock The {B}ehavior of {E}fficient {V}irtual {M}achine {I}nterpreters on
  {M}odern {A}rchitectures.
\newblock In {\em Proceedings of the 7th International Euro-Par Conference
  Manchester on Parallel Processing}, Euro-Par '01, pages 403--412, London, UK,
  2001. Springer-Verlag.

\bibitem{ErtlGregg03}
M.~A. Ertl and D.~Gregg.
\newblock The {S}tructure and {P}erformance of {E}fficient {I}nterpreters.
\newblock {\em The Journal of Instruction-Level Parallelism}, 5:1--25, Nov
  2003.

\bibitem{ErtlPirker97}
M.~A. Ertl and C.~Pirker.
\newblock The {S}tructure of a {Forth} {N}ative {C}ode {C}ompiler.
\newblock In {\em EuroForth '97 Conference Proceedings}, pages 107--116,
  Oxford, 1997.

\bibitem{ErtlPirker98}
M.~A. Ertl and C.~Pirker.
\newblock Compilation of {S}tack-{B}ased {L}anguages.
\newblock Final report to {FWF} for research project {P11231}, Institut f{\"u}r
  Computersprachen, Technische Universit{\"a}t Wien, 1998.

\bibitem{Ierusalimschy07}
R.~Ierusalimschy, L.~H. de~Figueiredo, and W.~Celes.
\newblock The evolution of lua.
\newblock In {\em HOPL III: {P}roceedings of the 3rd {ACM} {SIGPLAN}
  {C}onference on {H}istory of {P}rograming {L}anguages}. ACM Press, 2007.

\bibitem{Ierusalimschy96}
R.~Ierusalimschy, L.~H. de~Figueiredo, L.~Henrique, F.~Waldemar, and W.~C.
  Filho.
\newblock Lua - an {E}xtensible {E}xtension {L}anguage, 1996.

\bibitem{Intel10Vol1}
Intel Corporation.
\newblock {\em Intel 64 and IA-32 Architectures Software Developer's Manual
  Volume 1: Basic Architecture}, June 2010.
\newblock \url{http://www.intel.com/products/processor/manuals/}.

\bibitem{Kobalicek10}
P.~Kobalicek.
\newblock {A}sm{J}it.
\newblock \url{http://code.google.com/p/asmjit/}, 2010.

\bibitem{Lattner02}
C.~Lattner.
\newblock {LLVM}: {A}n {I}nfrastructure for {M}ulti-{S}tage {O}ptimization.
\newblock Master's thesis, Computer Science Dept., University of Illinois at
  Urbana-Champaign, Urbana, IL, Dec 2002.

\bibitem{Lattner04}
C.~Lattner and V.~Adve.
\newblock {LLVM}: {A} {C}ompilation {F}ramework for {L}ifelong {P}rogram
  {A}nalysis \& {T}ransformation.
\newblock In {\em Proceedings of the 2004 International Symposium on Code
  Generation and Optimization (CGO'04)}, Palo Alto, California, Mar 2004.

\bibitem{Leroy90}
X.~Leroy.
\newblock The {ZINC} {E}xperiment: {A}n {E}conomical {I}mplementation of the
  {ML} {L}anguage.
\newblock Technical Report 117, INRIA, Feb 1990.

\bibitem{Leroy10}
X.~Leroy, D.~Doligez, A.~Frisch, J.~Garrigue, D.~Remy, J.~Vouillon, et~al.
\newblock The {OCAML} language (version 3.12.0).
\newblock \url{http://caml.inria.fr/ocaml/}, 2010.

\bibitem{Leroy02}
X.~Leroy et~al.
\newblock The {C}aml {L}ight language (version 0.75).
\newblock \url{http://caml.inria.fr/caml-light/}, 2002.

\bibitem{LindholmYellin99}
T.~Lindholm and F.~Yellin.
\newblock {\em {T}he {Java\texttrademark} {V}irtual {M}achine {S}pecification}.
\newblock Prentice Hall PTR, 2nd edition, April 1999.

\bibitem{Matz10}
M.~Matz, J.~Hubicka, A.~Jaeger, and M.~Mitchell.
\newblock {\em System V Application Binary Interface: {AMD64} Architecture
  Processor Supplement (Draft Version 0.99.5)}, Sep 2010.
\newblock \url{http://www.x86-64.org/documentation.html}.

\bibitem{Pall10}
M.~Pall.
\newblock Lua{JIT}, 2010.
\newblock \url{http://luajit.org/}.

\bibitem{Remy02}
D.~Remy.
\newblock Using, {U}nderstanding, and {U}nraveling the {OCaml} {L}anguage
  {F}rom {T}heory to {P}ractice and {V}ice {V}ersa.
\newblock In G.~Barthe et~al., editors, {\em Applied Semantics}, volume 2395 of
  {\em Lecture Notes in Computer Science}, pages 413--537, Berlin, 2002.
  Springer-Verlag.

\bibitem{Starynkevitch04}
B.~Starynkevitch.
\newblock {OCAMLJIT} a faster {J}ust-{I}n-{T}ime {O}caml {I}mplementation,
  2004.

\bibitem{Taha06}
W.~Taha, C.~Calcagno, X.~Leroy, E.~Pizzi, E.~Pasalic, J.~L. Eckhardt,
  R.~Kaiabachev, O.~Kiselyov, et~al.
\newblock {MetaOCaml} - {A} compiled, type-safe, multi-stage programming
  language, 2006.
\newblock \url{http://www.metaocaml.org/}.

\bibitem{Vaswani03}
K.~Vaswani and Y.~N. Srikant.
\newblock Dynamic recompilation and profile-guided optimisations for a {.NET}
  {JIT} compiler.
\newblock {\em IEEE Proceedings - Software}, 150(5):296--302, 2003.

\end{thebibliography}

\end{document}